\documentstyle[prb,aps,twocolumn]{revtex}
\begin{document}
\draft
\title{Coherent dynamics of localized spins coupled with a two-dimensional hole gas
in diluted-magnetic quantum wells}
\author{K.V.Kavokin}
\address{{\it A. F. Ioffe Physico-Technical Institute, 194021 Politechnicheskaya 26, }%
St. Petersburg, Russia}
\date{ }
\maketitle

\begin{abstract}
Spin dynamics in a quantum well made of diluted-magnetic semiconductor is
studied theoretically. The exchange interaction of the ensemble of localized
spins with two-dimensional heavy-hole gas is shown to affect time evolution
of the spin system in an in-plane magnetic field. The Larmor frequency of
localized spins is reduced under influence of an oscillating effective field
of holes.
\end{abstract}


\narrowtext

Nanostructures grown on the base of diluted-magnetic semiconductors (DMSs)
combine quantum confinement of band- edge electron states with spin
magnetism of electrons localized in deep shells of magnetic ions (usually Mn$%
^{2+}$ ) embedded in a II-V semiconductor matrix. Recently, pronounced
spin-dynamic effects in a magnetic field applied have been found in such
structures \cite{Ref.1,Ref.2,Ref.3,Ref.4,Ref.5,Ref.6,Ref.7}. The effects
have their origin in the exchange interaction of confined carriers with
magnetic- ion spins. Briefly, they are governed by a collective precession
of many spins of magnetic ions, induced by an effective exchange field of
photoexcited heavy holes. The direction of this field, normal to the
quantum-well (QW) plane, is stabilized by quantum confinement via spin-
orbit interaction in the valence band, while the external magnetic field is
applied along the plane. These findings offer a possibility of manipulating
spin states in nanometer-sized objects by precise and high-speed optical
techniques. Possible applications range from characterization of
low-dimensional semiconductor structures (an example can be already found in
Ref.2 ) to quantum computation \cite{Ref.8}.

While dynamical properties of tightly bound spin systems of magnetic ions
and {\it localized} carriers (or excitons) have been fairly well understood
in terms of magnetic-polaron states \cite{Ref.1,Ref.5,Ref.7,Comments},
time-domain studies of DMS nanostructures containing {\it free} carriers are
at very beginning. One would expect these objects to show up a variety of
bright physical phenomena owed to many-body effects. A steady-state
photoluminescence study of a modulation-doped DMS quantum well with
two-dimensional (2D) hole gas at temperature below 2 K has already evidenced
a transition to a ferromagnetic phase\cite{Ref.9}, induced by holes via RKKY
mechanism\cite{Kittel}. A comparison of theoretical approaches to the
problem \cite{Ref.10} has demonstrated validity of a rather simple
mean-field description of the ordered phase: An effective exchange field
produced by polarized holes induces an average magnetization in the ionic
spin system, that in turn supports the orientation of hole spins. Due to low
concentration of carriers, the spin structure of the ferromagnetic phase is
very different from that typical of good metals\cite{Kittel}. Namely, mean
spin of magnetic ions remains much smaller than its maximum value, while
magnetization of holes is saturated. Studying the dynamics of the spin
system near the phase transition would elucidate the roles of interactions
involved. Recently developed techniques of time-resolved Faraday- \cite
{Ref.11} and Kerr-\cite{Ref.12} rotation show promise as experimental
approaches to this problem. The system can be driven away from the
equilibrium by the exchange field of excess holes created by a
circular-polarized light pulse, and then probed by another,
linear-polarized, pulse, as it has been done earlier for quantum wells
containing no equilibrium carriers \cite{Ref.3,Ref.6}. The rotation angle of
the polarization plane of the probe pulse is proportional to the component
of magnetization along the structure axis. An advantage of the method is
that the state of the spin system can be traced long after the excess
carriers created by the exciting pulse have recombined.

In this paper, the dynamics of the coupled spin systems of 2D hole gas and
magnetic ions in a quantum well subjected to an in-plane magnetic field is
analyzed within a mean-field approach. It is shown that interaction of
magnetic ions with the hole gas results in a renormalization of their Larmor
frequency. The frequency decreases with lowering temperature, and approaches
zero near the transition into the ordered phase. In the ferromagnetic phase,
the frequency remains virtually unchanged.

In the rest of the paper, we will use the following notation: The axis Z is
directed normal to the quantum well plane; the direction of the external
magnetic field, ${\bf B}$, that is supposed to lie in the QW plane, is
chosen for the axis X. Magnetic fields are expressed in energy units, $\hbar
=1$.

Spin systems of carriers and magnetic ions in DMSs are governed by the
following spin Hamiltonian

\begin{equation}
\hat H_S=a\sum_{i,j}\left( {\bf \hat J}_j,{\bf \hat S}_i\right) \delta
\left( {\bf r}_j^C-{\bf r}_i^I\right) +\hat H_{SS}+\sum_i\left( {\bf B},{\bf 
\hat S}_i\right)  \label{e1}
\end{equation}

where $a$ is a parameter of the exchange interaction, ${\bf \hat J}_j$ and $%
{\bf \hat S}_i$ are operators of spins of the $j$-th carrier (for holes, $J$%
=3/2) and of $i$-th magnetic ion, respectively. ${\bf r}_j^C$and ${\bf r}_i^I
$ are their position vectors \cite{Furdyna,Gaj}. The first term in Eq.(\ref
{e1}) describes the carrier-ion exchange interaction. The second term, $\hat 
H_{SS}$, accounts for superexchange interaction of adjacent magnetic ions,
and the last one gives Zeeman energy of magnetic-ion spins in an external
magnetic field. Much smaller Zeeman energy of carriers is omitted. In
addition, holes are influenced by strong spin-orbit interaction usually
expressed in the form of the Luttinger Hamiltonian, $\hat H_L$\cite{Ref.13}.
Together with the quantum-well confinement, the interaction splits the hole
states near the top of the valence band into the light- and heavy- hole
subbands, both of them retaining two-fold Kramers degeneration. The spin
doublets are conveniently described within a pseudospin formalism first
introduced in the theory of magnetic resonances \cite{Abragam} and then
applied also to DMS QWs \cite{MK}. In the following, we will use a truncated
Hamiltonian, obtained in spherical approximation by expansion of the full
Hamiltonian of the system, $\hat H_S+$ $\hat H_L$, up to second powers of
hole wave vectors near the top of the heavy-hole subband, where the Kramers
doublet corresponds to $J_z=\pm 3/2$ . In terms of secondary quantization,
this Hamiltonian reads:

\begin{eqnarray}
\hat H_T &=&\sum_{{\bf k},\sigma }\frac{k^2}{2m_{hh}^{\bot }}\hat a_{{\bf k}%
,\sigma }^{+}\hat a_{{\bf k},\sigma }+\sum_iB\hat s_{0,i}^x+\hat H_{SS}+ 
\nonumber \\
&&\ +aJ\sum_i\left| \Psi _h\left( z_i\right) \right| ^2\sum_{{\bf k},{\bf q}%
,\sigma }\left( \hat s_{{\bf q},i}^z\hat \sigma _z\right) \hat a_{{\bf k}%
,\sigma }^{+}\hat a_{{\bf k+q},\sigma }+  \nonumber \\
&&\ +aJQ\frac{L^2}{\pi ^2}\sum_i\left| \Psi _h\left( z_i\right) \right|
^2\sum_{{\bf k},{\bf q},\sigma ,\sigma ^{\prime }}\hat a_{{\bf k},\sigma
}^{+}\hat a_{{\bf k+q},\sigma ^{\prime }}\times  \label{e2} \\
&&\ \times \left( \left( {\bf k},{\bf \hat s}_{{\bf q},i}^{\bot }\right)
\left( {\bf k+q},{\bf \hat \sigma }\right) +\left( {\bf k+q},{\bf \hat s}_{%
{\bf q},i}^{\bot }\right) \left( {\bf k},{\bf \hat \sigma }\right) -\right. 
\nonumber \\
&&-\left. \left( {\bf k},{\bf k+q}\right) \left( {\bf \hat s}_{{\bf q}%
,i}^{\bot },{\bf \hat \sigma }\right) \right)  \nonumber
\end{eqnarray}
Here ${\bf k}$ is the two-dimensional wave vector of a hole, $m_{hh}^{\bot }$
is the heavy-hole in-plane effective mass, ${\bf \hat s}_{{\bf q},i}$ is a
two- dimensional Fourier transform of the spin-density operator of the
magnetic ions in a monolayer with the coordinate $z_i$ , $\sigma $ denotes
the states of the heavy-hole spin doublet, $\hat \sigma _x,\hat \sigma _y,%
\hat \sigma _z$ are Pauli matrices related to pseudospin components. The
components of vectors perpendicular to $Z$ are marked with the sign ''$\bot $%
''. $\Psi _h(z)$ is the hole wave function determined by the confinement
potential of the QW, $L$ is the QW width. The numerical factor, $Q\sim 1$,
depends on the details of the QW energy spectrum and can be found for any
specific structure by use of the perturbation theory.

Further, we will concentrate on the evolution of spin density vectors of
ions and holes averaged over the QW area. $Z$-components of these vectors
contribute to the observed optical signals. Within a mean-field approach,
the carrier-ion interaction has to be represented in terms of exchange
fields, proportional to average spins of carriers and magnetic ions, that
affect magnetic-ion and carrier spins, respectively\cite{Ref.7,Comments,Gaj}%
. One can separate a mean- field Hamiltonian out of Eq.(\ref{e2}) by
retaining there only components of ${\bf \hat s}_i$ at ${\bf q}$=0 and
averaging over ${\bf k}$. This operation readily leads one to the conclusion
that neither holes, nor ions create in-plane components of mean fields
acting upon mean spins of each other. The mean fields of holes and ions, $%
{\bf b}_{hi}$ and ${\bf B}^I$ respectively, are given by the expressions :

\begin{eqnarray}
{\bf b}_{hi}\left( t\right) &=&a\left| \Psi _h\left( z_i\right) \right| ^2%
{\bf j}  \nonumber \\
{\bf B}^I &=&a\sum_i\left| \Psi _h\left( z_i\right) \right| ^2{\bf s}_i
\label{e3}
\end{eqnarray}
where ${\bf j}$ is the hole spin density (${\bf j}$ is parallel to Z), ${\bf %
s}_i$ are average values of operators ${\bf \hat s}_i$ . The mean spins
evolve under $Z$-components of this fields. The extreme anisotropy of the
exchange interaction of spins averaged over entire ensembles of holes and
magnetic ions plays the decisive role in their spin dynamics, as discussed
below.

The spin subsystems are also coupled by the last term in Eq.(\ref{e2}),
which brings about spin flips. While the relaxation of ionic spins is most
likely dominated by the non-scalar part of $\hat H_{SS}$ \cite{Ref.14}, one
can suppose the interaction with the magnetic-ion subsystem to be the main
reason for rapid spin relaxation of holes, within less than 1ps, reported in
diluted-magnetic QWs \cite{Ref.6}. It should be even more effective for
Fermi-edge holes in modulation-doped structures, as they have considerable
wave vectors even at low temperature (note that the matrix element is
proportional to $k^2$). The detailed theory of hole spin relaxation is
beyond the scope of this paper; yet the fact that the Hamiltonian Eq(\ref{e2}%
) allows for such relaxation processes, together with experimentally
documented short spin-decay times, is of importance for further analysis of
the{\it \ coherent} spin dynamics in the system.

Let us now consider the dynamics of the mean spin density of magnetic ions,
given by the vectors ${\bf s}_i$. They evolve under the external magnetic
field, ${\bf B}$, and effective exchange fields, ${\bf b}_{hi}$, produced by
holes at each monolayer. It is worth to note that, at the conditions of the
experiment \cite{Ref.6}, the strength of the external field $B$ is of the
order of a few Tesla, while even completely polarized hole gas with area
concentration $2\cdot 10^{11}cm^{-2}$ , the value typical of
modulation-doped structures, can create an exchange field of less than 0.1T 
\cite{Ref.9}. Therefore, the vectors ${\bf s}_i$ remain directed very close
to the direction of ${\bf B}$ during the evolution of the spin system. At
these conditions, time evolution of each ${\bf s}_{i\text{ }}$is described
by just two differential equations for its $Z$- and $Y$-components, while
the value of $s_x$ is approximately equal to its equilibrium value in the
external field:

\begin{equation}
\left\{ 
\begin{array}{l}
\dot s_{iz}=-Bs_{iy}-\frac{s_{iz}-s_{0z}\left( b_{hi}\left( t\right) \right) 
}{T_2} \\ 
\dot s_{iy}=Bs_{iz}-b_{hi}\left( t\right) s_{ix}-\frac{s_{iy}}{T_2} \\ 
s_{ix}=s_{0x}\left( B\right)
\end{array}
\right.  \label{e4}
\end{equation}
Here $s_{0x}(B)$ and $s_{0z}(b_{hi})$ are equilibrium values of the
corresponding spin-density components in fields $B$ and $b_{hi}$,
respectively. $T_2$ is the transverse relaxation time of ionic spins.

As mentioned above, the interaction of holes with magnetic ions provides
fast spin relaxation of holes. The experimentally measured spin relaxation
time of holes, $\tau _{sh}<1ps$ \cite{Ref.6}, is much shorter than any
relaxation time in the magnetic-ion spin system ($>100ps)$, and than the
period of Larmor precession in actual magnetic fields (20$\div 100ps)$.
Thus, we can describe the dynamical interaction between spin subsystems of
magnetic ions and holes in terms of a time- dependent hole spin density
governed by the {\it instant} value of the magnetic-ion exchange field: $%
j(t)=j(B_z^I(t))$. Under this assumption, the system of equations, Eq.(\ref
{e4}), becomes closed. In order to solve the system, we multiply the
equations by $a\left| \Psi _h\left( z_i\right) \right| ^2$ and, after
summing over $i$, obtain a set of equations for components of ${\bf B}^I$:

\begin{equation}
\left\{ 
\begin{array}{l}
\dot B_z^I=-BB_y^I-\frac 1{T_2}\left( B_z^I-a^2Ws_{0x}\left( B\right)
j\left( B_z^I\right) /B\right) \\ 
\dot B_y^I=BB_z^I-a^2Ws_{0x}\left( B\right) j\left( B_z^I\right) -B_y^I/T_2
\end{array}
\right.  \label{e5}
\end{equation}
where $W=\sum\limits_i\left| \Psi _h\left( z_i\right) \right| ^4$. The
stationary states of the system (Eq.(\ref{e5})) correspond to $B_y^I=0$,
while $B_z^I$ is defined by the equation:

\begin{equation}
BB_z^I-a^2Ws_{0x}\left( B\right) j\left( B_z^I\right) =0  \label{e6}
\end{equation}

Eq.(\ref{e6}) has one solution, $B_z^I=0$, under the condition:

\begin{equation}
\zeta =a^2W\frac{s_{0x}\left( B\right) }B\left. \frac{dj\left( B_z^I\right) 
}{dB_z^I}\right| _{B_z^I=0}\leq 1  \label{e7}
\end{equation}

If this condition is not satisfied, there are three stationary states, the
state with $B_z^I=0$ being unstable. Two stable states are situated
symmetric with respect to $B_z^I=0$. Each of them corresponds to saturated
magnetization of holes and some nonzero $z$-component of the ionic spin
density. This bifurcation, which can be passed by changing temperature due
to temperature dependence of $s_{0x}\left( B\right) /B$ \cite{Oseroff}, is
nothing else than the carrier-induced ferromagnetic transition, studied
earlier\cite{Ref.9,Ref.10} for the case of zero in-plane magnetic field. It
is analogous also to the spontaneous symmetry break in formation of the
two-dimensional magnetic polaron \cite{MK,Ref.17}

Let us consider small oscillations of the spin system about stationary
states on both sides of the transition. They are described by Eqs.(\ref{e5})
linearized near the corresponding stationary state. If the condition Eq.(\ref
{e7}) is satisfied, then the eigenfrequency of the oscillations of ${\bf B}%
_I $ is given by the expression:

\begin{equation}
\omega =\sqrt{B^2\left( 1-\zeta \right) -\zeta ^2/4T_2^2}\approx B\sqrt{%
1-\zeta }  \label{e8}
\end{equation}
( we assume that $B\gg 1/T_2$). The damping time is also renormalized, being 
$\tilde T_2=T_2/(1-\zeta /2)$ instead of $T_2$.

Having found ${\bf B}^I(t)$ under appropriate initial conditions, one can
obtain from Eq.(\ref{e3}) time-dependent exchange fields $b_{hi}(t)$,
produced by holes at ions of each monolayer. Then, time-dependent ion spin
densities can be calculated using Eqs.(\ref{e4}), where $b_{hi}(t)$,
considered as explicit functions of time, play the role of driving forces.
In optical experiments , the oscillations are induced by a pulse of the
exchange field of photoexcited holes\cite{Ref.6}. In this case, the initial
conditions read $s_{iy}=0$, $s_{iz}=0$ , and the driving force has the form $%
b_{hi}\left( t\right) =a\left| \Psi _h\left( z_i\right) \right| ^2\left(
j_0e^{-t/\tau _{sh}}+j\left( B_z^I\left( t\right) \right) \right) $, where $%
j_0$ is the spin density of holes injected by the exciting light pulse. Due
to linearity of Eqs.(\ref{e4}), a solution to them at these conditions takes
the following general form: $s_{iz}\left( t\right) =\left| \Psi _h\left(
z_i\right) \right| ^2f_z\left( t\right) $, $s_{iy}\left( t\right) =\left|
\Psi _h\left( z_i\right) \right| ^2f_y\left( t\right) $, where $f_z\left(
t\right) $ and $f_y\left( t\right) $ do not depend on the number of the
monolayer. By summing these expressions, weighted by $\left| \Psi _h\left(
z_i\right) \right| ^2$, over the monolayers and comparing with Eq.(\ref{e3}%
), we get $f_z\left( t\right) =B_z^I\left( t\right) \left( aW\right) ^{-1}$, 
$f_y\left( t\right) =$. $B_y^I\left( t\right) \left( aW\right) ^{-1}$.
Solving Eqs(\ref{e5}) gives for a time $t\gg \tau _{sh}$:

\begin{eqnarray}
B_z^I\left( t\right) &=&\frac{s_{0x}\left( B\right) a^2Wj_0\tau _{sh}}{\sqrt{%
1-\zeta }}e^{-t/\tilde T_2}\sin \omega t  \nonumber  \label{e9} \\
B_y^I\left( t\right) &=&s_{0x}\left( B\right) a^2Wj_0\tau _{sh}e^{-t/\tilde T%
_2}\cos \omega t  \label{e9}
\end{eqnarray}

where $\omega $ is given by Eq.(\ref{e8}). Respectively,

\begin{eqnarray}
{\bf s}_i\left( t\right) &=&s_{0x}\left( B\right) \left( {\bf e}_x+aj_0\tau
_{sh}\left| \Psi _h\left( z_i\right) \right| ^2e^{-t/\tilde T_2}\times
\right.  \label{e10} \\
&&\times \left. \left( {\bf e}_z\left( 1-\zeta \right) ^{-1/2}\sin \omega t+%
{\bf e}_y\cos \omega t\right) \right)  \nonumber
\end{eqnarray}
where ${\bf e}_x$, ${\bf e}_y$ ,and ${\bf e}_z$ are unit vectors along
corresponding axes. Thus, the ends of spin-density vectors, ${\bf s}_i$ ,
move synchronously along similar orbits. The time-dependent angle of
Faraday- or Kerr rotation at these conditions will oscillate at the
frequency $\omega $, the oscillations damping out with the time constant $%
\tilde T_2$. One can see from Eq.(\ref{e8}) that $\omega $ is always less
than the precession frequency of free magnetic-ion spins, equal to $B$. It
turns to zero near the transition point corresponding to $\zeta =1$ (the
mean-field theory is not of course valid in the immediate vicinity of this
point). While $\zeta $ further rises beyond the phase transition, the spin
polarization of holes rapidly saturates, and their susceptibility, $dj\left(
B_z^I\right) /dB_z^I$, approaches zero. Therefore, the frequency of small
oscillations in the ferromagnetic phase remains close to $B$.

It is worth emphasizing a radical difference between this frequency shift
and the Knight shift of NMR frequency due to interaction of nuclear spins
with electrons. The Knight shift can be also considered in terms of an
effective electron field acting upon local spins. However, in our case the
effective field of holes is perpendicular to the external magnetic field,
and its time average is zero. The origin of the effect is entirely
dynamical. Its physical background is as follows. When the mean spin of
magnetic ions deflects from the QW plane and respectively from the direction
of the external field, its Zeeman energy increases as the second power of
the deflection angle. Since $Y$ and $Z$ components of a large spin directed
along $X$ are classically conjugated variables, the Larmor precession can be
considered as an oscillation in this parabolic potential \cite{Ref.5}. In
the presence of holes, the energy shift is reduced at the expense of the
energy of exchange interaction between normal-to-plane components of hole
and ion spins. This energy is also quadratic in the deflection angle but
negative. Therefore, the interaction with holes results in lowering the
rigidity of the system (which is proportional to 1-$\zeta $), and its
eigenfrequency respectively decreases. Such a softening of oscillation modes
is typical of systems undergoing phase transitions.

Let us now outline the scope for experimentation, offered by the above
theory. First of all, it should be mentioned that the discussed cooperative
effect might be the reason for a shift of the Larmor frequency already
observed \cite{Ref.6} at short delay times after the exciting pulse, while
some residual {\it non-equilibrium} carriers were present in the QW. As the
concentration of holes in that experiment was essentially time-dependent,
and carriers might be overheated due to high power of the pump pulse, we
refrain from attempts to describe the results of Ref\cite{Ref.6}
quantitatively. Experiments performed on modulation-doped structures
containing equilibrium holes, or on undoped structures under controllable
constant-wave illumination, would provide more feasible and detailed
information on the behavior of the spin system. The ferromagnetic transition
can be detected by the change of the precession frequency near the
transition point, discussed above. For the ferromagnetic phase, the
suggested method shows promise of getting information on the spin order,
which is still lacking \cite{Ref.9}. For instance, a signal at some
different frequency would come from domain walls if any domain structure
exists. In the paramagnetic state, measurement of the precession frequency
can yield, by using Eqs.(\ref{e7}) and (\ref{e8}), the paramagnetic
susceptibility of hole gas, $dj/dB_z^I$, at various hole concentrations.
This quantity, equal to $9m_{hh}^{\bot }/4\pi $ for a 2D Fermi-gas of
non-interacting heavy holes at low temperature, is known to be modified by
the Coulomb interaction (for a review, see \cite{Ando}), which has been
suggested\cite{Ref.9,Ref.10} as one of the necessary conditions for
observation of the phase transition. For example, the fit of experimental
data\cite{Ref.9} implies that, for the 80\AA -thick Cd$_{0.975}$Mn$_{0.025}$%
Te QW with CdMgZnTe barriers, $\zeta =1$ at 1.8K (the transition point),
while a calculation for the case of a perfect hole gas gives $\zeta =$0.4.
It is noteworthy that at actual areal concentrations of holes, $p\sim
10^{11}cm^{-2}$, the dimensionless interaction parameter $r_S$ \cite{Ando}
is not small. This casts doubt on perturbative calculations of $dj/dB_z^I$ .
Under these circumstances, the suggested experimental approach, suitable for
low concentrations of holes, may be valuable for understanding spin
properties of 2D fermion systems.

In summary, we have studied dynamical properties of spin systems of magnetic
ions and hole gas, coupled by the exchange interaction in diluted-magnetic
quantum wells. The extreme spin anisotropy of two-dimensional heavy holes
results in a peculiar collective excitation that shows up as oscillations of
components of the ionic spin density in a magnetic field. The eigenfrequency
of this oscillation mode is reduced with respect to the Larmor frequency of
magnetic ions. It turns to zero near the carrier-induced ferromagnetic
transition. The phenomenon can be observed in time domain by means of
pump-and-probe polarization spectroscopies.

The author is grateful to V.L.Korenev, I.A.Merkulov, D.Scalbert,
M.N.Tkachuk, and D.R.Yakovlev for helpful discussions. The work was
supported in part by the Russian Foundation for Basic Research (Grant No
96-16941), Volkswagen Foundation, and DFG (SFB 410).

\end{document}